# Dispersion of deterministic sound wave into stochastic noise by a linear acoustic layer with time-varying and quantized material properties

*Yumin Zhang, Keming Wu, Chunqi Wang and Lixi Huang*[*]

Dr. Y. Zhang, K. Wu, Dr. C. Wang, Prof. L. Huang

Department of Mechanical Engineering, The University of Hong Kong

Pokfulam, Hong Kong, 00000, Hong Kong SAR

E-mail lixi.huang@hku.hk

Dr. Y. Zhang, K. Wu, Dr. C. Wang, Prof. L. Huang

Lab for Aerodynamics and Acoustics, Zhejiang Institute of Research and Innovation,

The University of Hong Kong , 1623 Dayuan Road, Hangzhou, Zhejiang 311300,

China

Abstract:

Wave is crucial to acquiring information from the world and its interaction with matter is determined by the wavelength, or frequency. Search for the ability to shift frequency often points to material nonlinearity, which is significant only when the excitation level is very high. Temporal modulated material is expected to shift sound frequency in a linear manner, but is so far not effectively realized due to inadequate modulation ratio. This study introduces a new class of temporal modulation material with a giant modulation ratio. When the modulation is given in a random time sequence, a monochromatic sound wave is converted to a white noise of a continuous frequency band. The demonstrated device is called a randomized acoustic meta-layer, consisting of a suspended diaphragm shunted by an analog circuit. The circuit alters the acoustic impedance of the layer which operates in two quantized states when the shunt is connected and disconnected by a MOSFET with a pseudo-random time sequence. The device has unprecedented potential applications such as converting annoying tones to white-noise-like hum to improve psychoacoustic quality, allowing encrypted underwater communication, and super-resolution imaging.

# 1. Introduction

Wavelength is the dominant parameter for manipulating a wave or using it to acquire information. The ability to change the time-base of a wave, or frequency, is highly desirable. Conventional materials, including spatial modulation metamaterials such as photonic and sonic crystals[1-6], have time-invariant physical properties. The search for time-varying or dynamic materials[7] often points to material nonlinearity, such as nonlinear optical crystal[8] and ultrasound contrast agent[9]. It creates new signal frequencies by harmonic wave generation and offers super-resolution in optical[10-12] and ultrasonic[9, 13-14] imaging, rectification and reciprocity breaking[15-18], and directional radiation[19-21]. However, such nonlinear effects require a high level of excitation not readily applicable in daily life. For example, to obtain a giant nonlinearity in air excited by ultrasound, it requires a local Mach number close to unity, or a wave amplitude of 146 kPa[9], much higher than the pain threshold of 63 Pa for human ears in the audible frequency range.

Linear temporal modulation mechanism can convert the frequency of electrical signal, which can be traced back to the amplitude modulation (AM) method in the telecommunication system[22-23]. AM utilizes time-varying circuit to mount the signal onto the carrier to produce a long-distance transmittable wave. The transmitted signal contains the difference and sum frequencies of the signal and the carrier. Recent research in temporal modulation material[24-34] shares the same core spirit with AM, and is successful in breaking reciprocity in a linear manner[25-32]. Existing temporal modulation of materials include two mechanisms, moving medium (biasing) and wave modulation[34]. Biasing is like the Doppler effect[26-27]. It requires a very high motion speed or momentum to obtain significant frequency shift which makes applications difficult. The wave modulation, on the other hand, changes the local medium properties, such as impedance, in a space-time sequence similar with a travelling wave[29-33]. However, the modulation ratio, which is defined as time-dependent magnitude to the DC magnitude of the time-varying parameter, is insufficient for existing temporal materials to achieve meaningful conversion of sound wave frequency.

In this work, a linear acoustic meta-layer with a giant modulation ratio is introduced, which realizes two types of sound wave frequency conversions. The first is converting the color, or rather frequency in acoustics, of a monochromatic sound wave to another color with unprecedented efficiency, and the second is dispersing a monochromatic sound to white noise of any prescribed frequency band. The dynamic response of the second meta-layer is random in time. For the first time, a material has general randomness and the mechanism is here called a randomized acoustic meta-layer (RAML) contrasting to the conventional static and dynamic materials which are governed by deterministic dynamic equations. These attributes have many potential applications. First, random modulation dilutes the time-signature of the incident sound, rendering its echo undetectable. Second, the broadband modulation widens the frequency scope of energy pumping and can be a new building block for parametric gain medium[35-37] for sound and offer better ultrasound and underwater imaging. Third, random material may become a new tool in environment noise control. For instance, tonal noise found in our living, working and hospital environments is highly annoying and can be a serious health issue[38]. Clinical trials show that random noise (white or pink) promotes sleep for the neonates[39], patients in intensive care unit[40] and coronary care unit[41].

## 2. The meta-layer architecture and theory

The proposed RAML is shown in **Figure 1(a)**. It consists of a suspended diaphragm of a moving-coil loudspeaker, shunted by a circuit. The coil is immersed in a permanent magnetic field and gives an electromechanical coupling of the series shunt circuit and diaphragm. Our previous work [42-43] demonstrates that the passive shunt circuit can rapidly and significantly alter the system dynamic properties, including mass, stiffness and damping. When shunt circuit is connected, the electromagnetically induced acoustic impedance is[42] $\Delta Z = (Bl)^2 Z_e^{-1}$, where $Z_e = R + Li\omega + (Ci\omega)^{-1}$ is the circuit electrical impedance with net resistance $R$, inductance $L$, capacitance $C$ and force factor $Bl$, and $\omega$ is the angular frequency. In this work, a metal-oxide-semiconductor field-effect transistor (MOSFET) is introduced to connect or cut off the shunt circuit. When the G-terminal of the MOSFET is applied with a bias voltage exceeding a threshold, $V_g > V_0$, the resistance between D and S is 4 m$\Omega$, and the shunt impedance $\Delta Z$ is loaded. Otherwise the resistance is $R_{off} = 4400$ $\Omega$, and $\Delta Z$ is unloaded. A state function $g(t) = H(V_g(t) - V_0)$ describes the MOSFET states, where H is the Heaviside step function. Figure 1(b) illustrates $g(t)$ when a periodical and random gating voltage $V_g(t)$ is used, respectively. Figures 1(c) and 1(d) illustrate, respectively, the working of RAML when $V_g(t)$ follows harmonic and random patterns. When the state of the MOSFET is controlled by a periodical voltage signal of frequency $f_m$, the energy of the incident sound wave at the source frequency of $f_s$ is dispersed to the difference and sum frequencies ($f_- = f_s-f_m$, and $f_+ = f_s+f_m$). When the gating voltage sequence is a band-limited random voltage with $f_m \in [f_1\ f_2]$, we expect the side frequencies to cover the same linear bandwidth of $f_2 - f_1$. The time sequence of the MOSFET is unrelated to the incident or transmitted waves, contrasting with traditional active control which is derived from a sensor signal and subject to stability constraints. In this sense, RAML is therefore a robust, passive wave scatterer which does not radiate sound on its own.

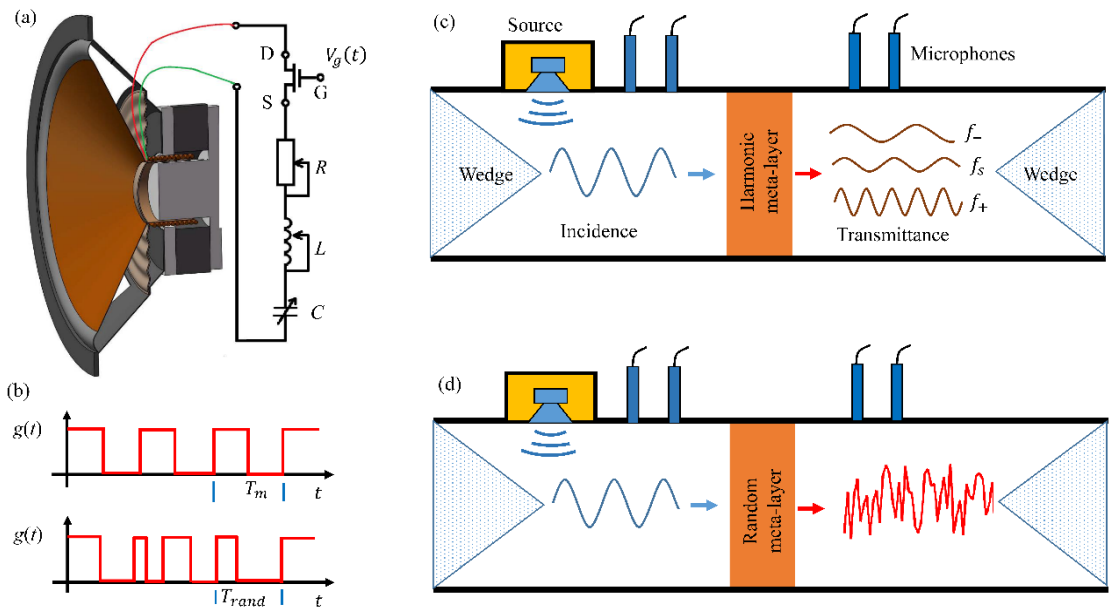

Figure1: Schematic of the proposed RAML and the conceptual diagram of frequency conversions. (a) RAML formed by a shunted loudspeaker cascading a MOSFET unit. (b) Schematic of the MOSFET state function, $g(t)$ =1 for "on", when the gating voltage $V_g$ >2 Volts [see (a)], and $g(t)$=0 for "off", otherwise. (c) Schematic of harmonic scattering in the waveguide. (d) Schematic of RAML with a random gating voltage, converting part of a tonal input to white noise.

The time-domain lumped-parameter governing equations for the dynamics of the meta-layer are,

$$\left. \begin{array}{l} M\dot{v}(t) + (2\rho_0 c_0 A + D)v(t) + K \int v(t)\,dt + BlI(t) = 2p_I(t)A \\ L\dot{I}(t) + \left[R + R_{off}(1 - g(t))\right]I(t) + C^{-1} \int I(t)\,dt = Blv(t) \end{array} \right\} \quad (1)$$

where over-dot donates time-derivatives, $p_I(t)$ is the incident sound pressure on the diaphragm, $v(t)$ is the velocity response of the diaphragm, $I(t)$ is the electrical current in the shunt circuit, $M$, $D$ and $K$ are, respectively, the dynamic mass, damping and stiffness, $A$ is the cross-section of the waveguide, $\rho_0$ and $c_0$ are the air density and speed of sound, respectively. The damping term with $2\rho_0 c_0 A$ accounts for the fluid loading on the downstream and upstream sides of the diaphragm when the waveguide is terminated by anechoic ends[44]. Applying Fourier transform to both sides of Equations (1) yields:

$$\left. \begin{array}{l} i\omega M\hat{v} + (2\rho_0 c_0 A + D)\hat{v} + \frac{1}{i\omega}K\hat{v} + Bl\hat{I} = 2A\hat{p}_I \\ i\omega L\hat{I} + \left(R + \frac{1}{2}R_{off}\right)\hat{I} - \frac{1}{2}R_{off}\hat{I} \otimes \hat{G} + (i\omega C)^{-1}\hat{I} = Bl\hat{v} \end{array} \right\} \quad (2)$$

where over-hats denote Fourier transform, the convolution $\hat{I} \otimes \hat{G}$ is calculated as $\int_{-\infty}^{\infty} \hat{G}(\omega')\hat{I}(\omega - \omega')d\omega'$, and $\hat{G}(\omega)$ is the Fourier transform of $2g(t)-1$ (square wave with 0 bias).

The presence of the convolution term, $\hat{I} \otimes \hat{G}$, is the essence of the modulation device. It makes the solution of Equations (2) complicated as responses at one frequency are coupled to the material properties at all other frequencies. However, preliminary analysis is still possible for simplistic modulations. For a periodic square wave $2g(t)-1$, with period $2\pi/\omega_m$, $\hat{G}(\omega)$ takes the sinc form, $\hat{G}(\omega) = \sum_{n=-\infty}^{\infty} \delta[\omega - (2n-1)\omega_m]/(2n-1)$, where $\delta$ is Dirac delta function. Note that the convolution $\hat{I} \otimes \hat{G}$ shifts the frequency of the electrical current $\hat{I}$ by $(2n-1)\omega_m$. The amplitude factor of 1/(2*n*-1) guarantees the dominance of the lowest (fundamental) orders of *n*=0,1 and this is confirmed by experimental results presented below.

Two factors in $\hat{I} \otimes \hat{G}$ determine the extent of frequency spread by the modulation. First, when $g(t)$ follows a completely random time sequence, $\hat{G}(\omega)$ becomes a constant for all frequencies. Second, the finite jump of the meta-layer impedance *ΔZ* between the shunt-off and shunt-on states (over nano seconds by MOSFET) makes the spectrum of $\hat{I}$ a mathematical constant. The meta-layer may be said to operate on two "quantized" impedance states. Together with the randomized switching, RAML offers a giant modulation ratio (see below). Following these frequency-domain observations, we now return to the time-domain governing Equations (1) which are numerically solved for the velocity response *v*(*t*). The transmitted and reflected sound pressures are [44],

$$p_T(t) = \rho_0 c_0 v(t), \; p_R(t) = p_I(t) - p_T(t). \tag{3}$$

To demonstrate the basic characteristics of the proposed modulation mechanism, harmonic modulation is examined before random modulation is applied on the fully functional RAML device.

**3. Experiment results**

**3.1. Results for the harmonic modulation**

We first scrutinize the static acoustic impedance for the meta-layer at MOSFET-on and MOSFET-off states using the impedance tube illustrated in Figure 1(c), while the results are shown in **Figure 2(a)**. The resonance frequency of the diaphragm itself, with MOSFET at "off" state, is 140 Hz, at which the static impedance of the diaphragm is 59% of the air impedance, $\rho_0 c_0 A$, while the magnitude of impedance for the MOSFET-on state is $35\rho_0 c_0 A$, dominated by the shunt induced damping. It means that, between the MOSFET-on and -off states, the impedance modulation ratio is $\alpha_m$= 35/0.59-1=58. This is at least two orders of magnitude higher than the modulation ratio deployed by the pioneering techniques in literature: 0.14~0.21 for vibration and sound[32, 33], and $10^{-4}$~$10^{-3}$ in optics[24, 25]. The modulation mechanism for RAML is therefore described as a "giant" modulation, which will be desirable in future applications described in the Introduction, and indeed beyond. For instance, one technique of achieving time-reversal-based holography is to create an instantaneous time mirror[45] by suddenly changing the global wave speed leading to back propagation of waves without using an antenna array on an enclosing boundary. A larger time disruption in wave speed, or modulation in the current context, will give more substantial back propagation.

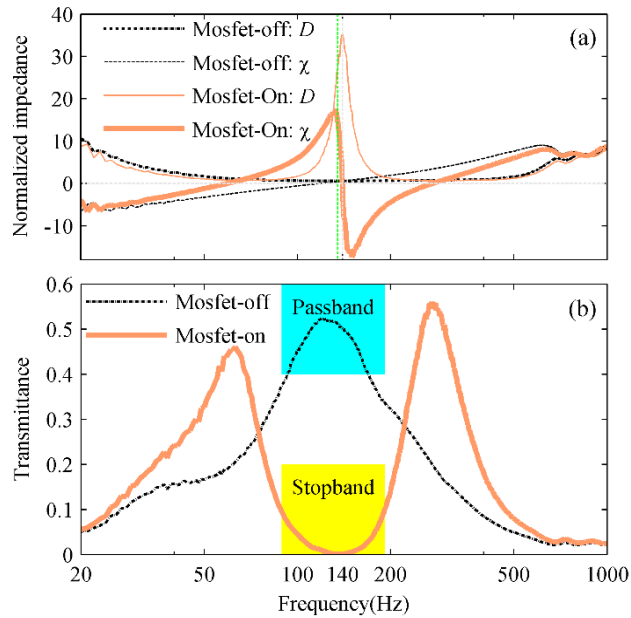

Figure 2: Results of the acoustic impedance measurement. (a) Acoustic impedance of the meta-layer, $D+i\chi$, for MOSFET-on and -off states. Both damping ($D$) and reactance

($\chi$) are normalized by the air impedance. The diaphragm properties are $M$ =5.8 g, $D$ =5.04 Nsm$^{-1}$, $K$ =4516 Nm$^{-1}$, with force factor $Bl$ =4.6 Tm. The circuit has $R$ =0.1 Ω, $L$ =1.2 mH and $C$ =1.0 mF (b) Sound transmittances showing passband and stopband at MOSFET off- and on- states, respectively.

Figure 2(b) shows, when the MOSFET is switched on, the meta-layer changes from an acoustic soft state to an acoustic rigid state within 100 Hz -200 Hz, therefore, the sound passband due to structural resonance is transformed to a stopband due to the extremely large damping induced by the shunt circuit. It acts like an instant phase-change material. Therefore, when the 'on' and 'off' states of the MOSFET oscillate with time, the incident sound wave is scattered in time history. Note that the supplied gating voltage does not provide energy for such virtual material change, nor does the meta-layer radiate sound on its own.

The results for the harmonic modulation are shown in **Figure 3** for an incident sound wave of 5.5 Pa at $f_s$ =135 Hz. The modulation frequency is $f_m$ =75 Hz. Figure 3(a) shows that the transmitted wave is distorted by the time-varying impedance of the meta-layer. The solid line is the measured signal, while the dashed line, labeled "reconstructed", is the sum of three frequency components, $f_s$, $f_-$, $f_+$. The close agreement between the two means that contributions from higher-order terms of modulation are negligible. Figure 3(b) shows the measured waveforms of the three frequency components separately. Their amplitudes, normalized by the incident wave, are 0.279, 0.184 and 0.186 for $f_s$, $f_-$, $f_+$, respectively. Figure 3(c) compares the measured amplitudes with the numerical prediction in the spectral form, showing reasonable agreements. The time-domain comparison is made in Figure 3(d) for the total transmitted sound.

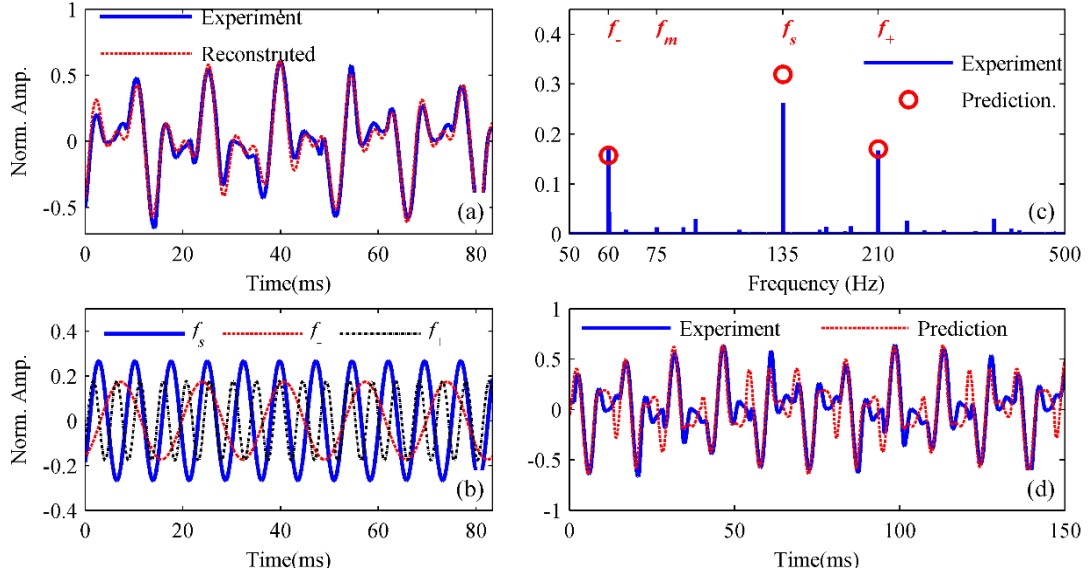

Figure 3: Measurement and prediction of harmonic modulation meta-layer. All amplitudes in the sub-figures are normalized by the incident wave amplitude. (a) Measured time-domain sound pressure (solid curve) and the sum (dashed) of the three frequency components of $f_s$, $f_-$ and $f_+$. (b) Decomposition into the source frequency $f_s$ (solid, 5.5Pa), the difference frequency $f_-$ (60Hz, 0.95 Pa, dashed) and the sum frequency $f_+$ (210Hz, 0.97 Pa, dot-dashed). The corresponding amplitudes for the reflected waves are 3.81 Pa, 1.05 Pa and 0.88 Pa for $f_s$, $f_-$ and $f_+$, respectively. (c) Spectra of the measured (solid) and numerically predicted (open circle) downstream pressures. (d) Time-domain prediction (dashed) of the transmitted pressure compared with the measured (solid line).

### 3.2. Results with random modulation

The voltage sequence of random modulation is obtained by setting positive and negative values of a band limited white signal to be $V_g$ = 6 Volts ($g$ =1) and $V_g$ = 0 ($g$ =0). This is illustrated in **Figure 4(a)** as the solid trace on the top. The rest of Figure 4 describes the full data set of RAML with an incident tone set at $f_s$ =135 Hz, and the random modulation function $g(t)$ band-limited to $f_m \in$ [50, 100] Hz. The difference and sum frequencies fall in the range of $f_-$ =[35, 85] Hz, and $f_+$ =[185, 235] Hz, respectively.

Figure 4(a) illustrates the special design of time windows for the experimental study. The incident wave signal (dashed line) is introduced from time 0 to 30sec, while the gating voltage (solid line) is introduced from 15sec to 45sec, producing a total of four time segments, labeled as i-iv in the abscissa. Figure 4(b) is the actual trace of the transmitted wave covering the whole 60 seconds. Figure 4(c) is the zoom-in view of the time window marked in Figure 4(b) for the randomization onset around the 15th second. The incident sound has an amplitude of 4.56 Pa. Significant distortion by RAML occurs as soon as the modulation is activated. Figure 4(d) shows a typical chunk of signal around the 17th second, decomposed into the residual source frequency ($f_s$ = 135 Hz, 1.36 Pa) and the randomized sound pressure which has an amplitude of 1.19 Pa. The latter represents some 43.4% of the total transmitted sound energy. Considering the use of a single RAML unit, this percentage of energy transfer from tonal sound to broadband noise is deemed efficient.

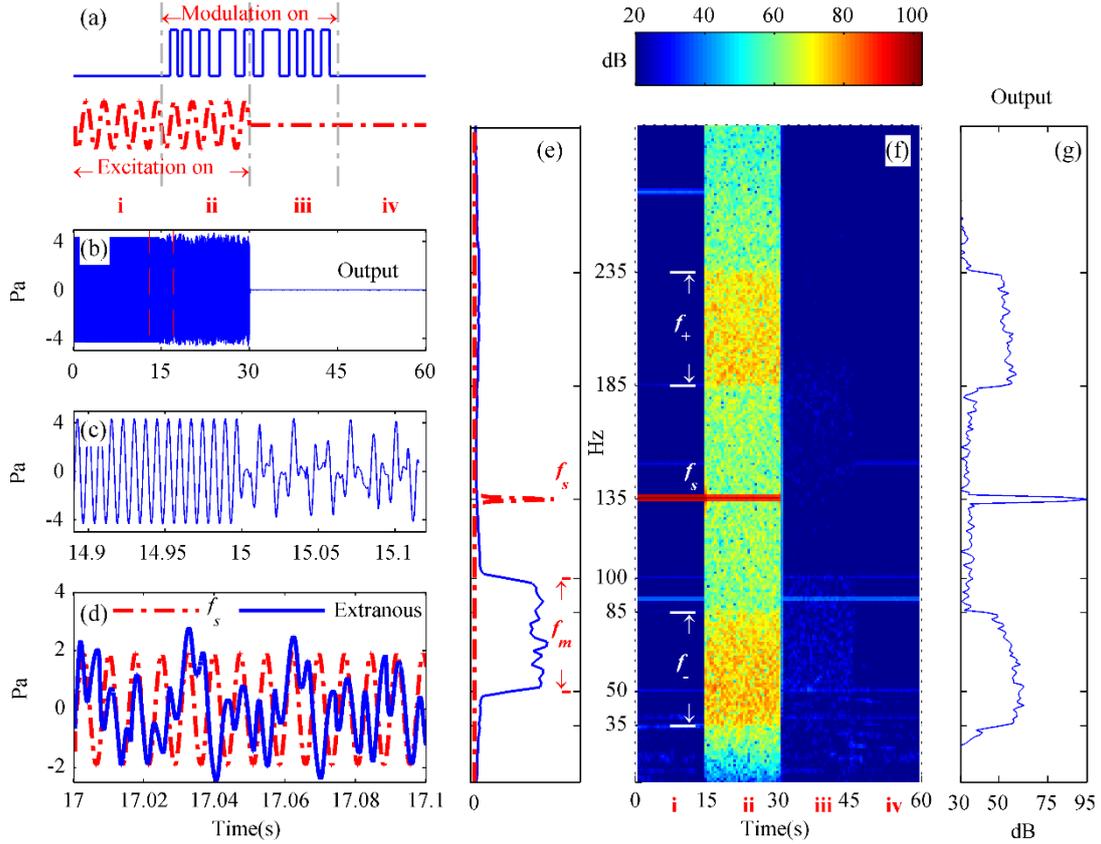

Figure 4: Experimental results for the conversion of monochromatic sound to band-limited white noise by RAML. (a) Illustration of the incident sound introduced from time 0-30 sec (signal period not to scale) and the random modulation introduced from 15- 45 sec, forming four time-segments labeled as i-iv. (b) Transmitted time-domain pressure signal. (c) Zoom–in-view of the transition from tonal sound to band-limited white noise around the 15$^{th}$ second. (d) Transmitted signal decomposed into the source frequency $f_s$ and randomized output from the 17$^{th}$ second. (e) Spectra of the generated driving signal (dot-dash, $f_s$) and the modulation voltage (solid line, $f_m$). (f) Spectrogram of the transmitted sound downstream, with frequencies of $f_s, f_-, f_+$ marked. (g) Spectrum of the transmitted pressure signal.

Figures 4(e)-4(g) share the same vertical coordinate for frequency. Figure 4(e)

shows the spectra of the incident tone (dashed line, labeled $f_s$) and the modulation signal (solid line, labeled $f_m$). Figure 4(f) is the spectrogram of the measured transmitted sound with color scale in the wide range of 20-100 dB. Sound pressures below 20 dB (0.2 mPa) are displayed by the color at 20 dB and treated as background noise. Time segment (i) features a single spectral peak at $f_s$ =135 Hz, as expected. Segment (ii) has somewhat weakened incident wave at frequency $f_s$, plus two finite bright bands of $f_-$ = [35, 85] Hz, and $f_+$ =[185, 235] Hz, which correspond clearly to the spectrum of the transmitted wave shown in Figure 4(g). Segment (iii) has modulation on but without incident sound. Zero output signal is expected here, but in reality there are gate leakage current of the MOSFET and voltage ripples caused by imperfect electrical ground lead, which causes the RAML to radiate a tiny amount of sound around 28 dB. Segment (iv) contains only the electronic noise in the measurement system.

With one RAML unit, the transmitted sound has around 56.6% of residual energy at the source frequency $f_s$. Cascading the device is expected to deplete the energy at the source frequency quickly and the final output cannot be detected for $f_s$ for a prescribed signal-to-noise ratio unless one has the precise modulation "key", $g(t)$. In this sense, the latter provides encryption for the incident sound wave, which can be useful for applications like underwater communication and avoidance of sonar detection. Another potential application is in ultrasound. Low frequency can penetrate the tissue deeply, but high frequency is needed for fine image resolution. If RAML is used as a back reflector, which creates a broadband in high frequency, it may cut the loss of high frequency wave energy by half compared with its direct emission at the sensor.

## 4. Conclusions

In summary, we take a broad view of the meta-layer operating at the MOSFET-on and MOSFET-off configurations, each with its own eigen spatial mode. When a step discontinuity occurs from one to the other, the difference between the two modes forms an impulse whose energy is scattered to all frequencies. This is a time-domain phenomenon, or a time crystal [32], similar with the space refraction of waves by a sharp boundary or diffraction around a singularity. If the MOSFET switching is repeated periodically, the time slab limits the scattering to frequencies correlated with the modulation period. The essence of the demonstrated RAML is the broadband frequency scattering taking place at a single MOSFET switch, while randomized switching prevents destructive interference between different time-scatter events. The key performance parameter is the modulation ratio. The giant ratio demonstrated here will be crucial for the success of potential applications, such as tonal noise dispersion, linear acoustic diode, encrypted underwater communication or misleading Doppler-based detection, parametric amplifier, super-resolution imaging and hologram, etc.

## 5. Experiment and calculation method

Negative impedance converter[42], is used for reducing the DC resistance of the coil to tune the effective resistance from 0 to desired positive values. Due to the reflection by the meta-layer, standing waves form in the upstream, which are decomposed to incident and reflection waves by signals collected by the 1$^{st}$ and 2$^{nd}$ microphones (GRAS, 8cm apart) following established procedures [46]. At downstream, two microphones are also used to decompose waves and ensure that the refection coefficient is indeed negligible.

The mechanical and electrical parameters of the diaphragm in Equations (1) are obtained by fitting the acoustic impedance of the diaphragm shown in Figure 2(b) for the frequency range of 30 Hz to 500 Hz. The amplitude of the incident wave is set as 5.5 Pa in the calculation. Equations (1) are solved for velocity and current responses, $v(t)$ and $I(t)$, in the time domain using the MATLAB algorithm of "ODE45". The transmitted and reflected waves are then obtained by Equations (3).

**Acknowledgements**

This work is supported by the National Science Foundation of China Project 51775467, the general research fund project 17210720 from the Research Grants Council of Hong Kong SAR, and a block grant from the Hangzhou Municipal Government. We also thank Miss Yating Liu for the artwork. Y.Z. and L.H. initiated the concept. Y.Z and K.W. designed and conducted the experiments. Y.Z. and C.W. carried out the numerical analysis. L.H., Y.Z. and K.W. wrote the manuscript, which is reviewed by all authors.

**Competing financial interests**

The authors declare no competing financial interests.

**References**

[1]    L. Brillouin, *Wave Propagation in Periodic Structures*, Dover Publications, New

York, USA, **1953**.

[2]     E. Yablonovitch, *Phys. Rev. Lett.* **1987**, *58*, 2059.

[3]     S. John, *Phys. Rev. Lett.* **1987,** *58*, 2486.

[4]     J. B. Pendry, *Phys. Rev. Lett.* **2000**, *85*, 3966.

[5]     R. A. Shelby, D. R. Smith, S. Schultz, *Science* **2011**, *292,* 77-79.

[6]     Z. Liu, X. Zhang, Y. Mao, Y. Y. Zhu, Z. Yang, C. T. Chan, P. Sheng, *Science*, **2000**, *289,* 1734-1736.

[7]     K. A. Lurie, *An Introduction to the Mathematical Theory of Dynamic Materials*, Springer, New York, USA **2007**.

[8]     R. W. Boyd, *Nonlinear Optics*, Academic press, Burlington, Canada, **2019**.

[9]     L. Hoff, *Acoustic Characterization of Contrast Agents for Medical Ultrasound Imaging*, Springer, Berlin, Germany **2001**.

[10]    Y. R. Shen, *The Principles of Nonlinear Optics*, Wiley, New York, USA, **1984**.

[11]    C. V. Raman, K. S. Krishnan, *Nature* **1928**, *121*, 501-502.

[12]    G. Turrell, J. Corset, *Raman microscopy: Developments and Applications*, Academic Press, San Deigo, USA **1996**.

[13]    M. A. Averkiou, D. N. Roundhill, J.E. Powers, 1997 IEEE Ultrason. Symp. Proc. Vol. 2, 1561-1566, IEEE, New York, USA, **1997**.

[14]    D. H. Simpson, C. T. Chin, P. N. Burns, *IEEE. T Ultrason. Ferr.* **1999**, *46*, 372-382.

[15]    B. Liang, X. S. Guo, J. Tu, D. Zhang, J. C. Cheng, *Nat. Mater.* **2010**, *9*, 989.

[16]    B. Liang, B. Yuan, J. C. Cheng, *Phys. Rev. Lett.* **2009**, *103*, 104301.

[17] N. Boechler, G. Theocharis, C. Daraio, *Nat. Mater.* **2011**, *10*, 665.

[18] B. I. Popa, S. A. Cummer, *Nat. Commun.* **2011**, *5*, 3398.

[19] P. J. Westervelt, *J. Acoust. Soc. Am.* **1963**, *35*, 535-537.

[20] M. Yoneyama, J. I. Fujimoto, Y. Kawamo, S. Sasabe, *J. Acoust. Soc. Am.* **1983**, *73*, 1532-1536.

[21] F. J. Pompei, *Sound from ultrasound: The parametric array as an audible sound source*, Ph.D. thesis, Massachusetts Institute of Technology, **2002**.

[22] L. De Forest (AT&T), U.S. 1507016, **1915**.

[23] L. De Forest (AT&T) U.S. 2126541, **1935**.

[24] S. Fan, M. F. Yanik, M. L. Povinelli, S. Sandhu, *Opt. Photonics News*, **2007**, *18*, 41-45.

[25] Z. Yu, S. Fan, *Nat. Photonics* **2009**, *3*, 91.

[26] D. L. Sounas, C. Caloz, A. Alu, *Nat. Commun.* **2013**, *4*, 2407.

[27] R. Fleury, D. L. Sounas, C. F. Sieck, M. R. Haberman, A. Alù, *Science* **2014**, *343*, 516-519.

[28] Y. Hadad, D. L. Sounas, A. Alu, *Phys. Rev. B* **2015**, *92*, 100304.

[29] N. Swinteck, S. Matsuo, K. Runge, J. O. Vasseur, P. Lucas, P. A. Deymier, J. Appl. Phys. **2015**, *118*, 063103.

[30] G. Trainiti, M. Ruzzene, *New J. Phys.* **2016**, *18*, 083047.

[31] J. Vila, R. K. Pal, M. Ruzzene, G. Trainiti, *J. Sound Vib.* **2017**, *406*, 363-377.

[32] Y. Wang, B. Yousefzadeh, H. Chen, H. Nassar, G. Huang, C. Daraio, *Phys. Rev. Lett.* **2018**, *121*, 194301.

[33] G. Trainiti, Y. Xia, J. Marconi, G. Cazzulani, A. Erturk, M. Ruzzene, *Phys. Rev. Lett.* **2019**, *122*, 124301.

[34] C. Caloz, Z. L. Deck-Léger, Part I-II, *IEEE Trans. Antennas Propag.* **2019**, *68*, 1569-1598.

[35] J. M. Manley, Some general properties of magnetic amplifiers. *IEEE Proc. IRE* **1951**, *39*, 242-251.

[36] P. K. Tien, *J. Appl. Phys.* **1958**, *29*, 1347-1357.

[37] J. C. Simon, *IEEE Trans. Microw. Theory Techn.* **1960**, *8*, 18-29.

[38] H. Xie, J. Kang, G. H. Mills, *Crit. Care* **2009**, *13*, 208.

[39] J. A. Spencer, D. J. Moran, A. Lee, D. Talbert, *Arch Dis. Child.* **1990**, *65*, 135-137.

[40] M. L. Stanchina, M. Abu-Hijleh, B. K. Chaudhry, C. C. Carlisle, R. P. Millman, *Sleep Med.* **2005**, *6*, 423-428.

[41] P. F. Afshar, F. Bahramnezhad, P. Asgari, M. Shiri, *J Caring Sci.* **2016**, *5*, 103.

[42] Y. Zhang, C. Wang, C., L. Huang, *Mech. Syst. Signal Proc.* **2019**, *126*, 536-552.

[43] Y. Zhang, C. Wang, L. Huang, *Appl. Phys. Lett.* **2020**, *116*, 183502.

[44] F. J. Fahy, P. Gardonio, *Sound and structural vibration: Radiation, Transmission and Response*, Elsevier, Oxford, England **2007**.

[45] V. Bacot, M. Labousse, A. Eddi, M. Fink, E. Fort, *Nat. Phys.* **2016**, *12*, 972-977.

[46] Y. J. Chung, D. A. Blaser, *J. Acoust. Soc. Am.* **1980**, *68*, 907.

[47]